\documentclass[12pt,letterpaper]{article}
\usepackage[usenames, dvipsnames]{xcolor}
\usepackage{jcapmod}

\usepackage{tocloft}
\usepackage[]{todonotes}
\usepackage{verbatim}
\usepackage{mathrsfs}
\usepackage{cleveref}
\usepackage[shortlabels]{enumitem}
\usepackage{pgf}
\usepackage{tikz-cd}
\usetikzlibrary{shapes,arrows}
\usetikzlibrary{calc}
\usepackage{pgfplots}
\pgfplotsset{compat=1.12}
\usepackage{tikz-3dplot}

\usepackage{amsthm}

\usepackage{tikz}
\usepackage{setspace,caption}
\usepackage{lipsum}
\usetikzlibrary{matrix,arrows,calc}
\makeatletter
\newsavebox\myboxA
\newsavebox\myboxB
\newlength\mylenA

\definecolor{cornellRed}{HTML}{B31B1B}
\definecolor{cornellBlue}{HTML}{0068AC}
\definecolor{cornellGreen}{HTML}{6EB43F}

\usepackage{bbm} 					
\usepackage{slashed} 				
\usepackage{graphicx}				
\usepackage{subcaption}			
\usepackage{psfrag}				
\usepackage{tensor}				
\usepackage{fouridx}				
\usepackage{bm}					
\usepackage{mdframed}				
\usepackage{multirow}				
\usepackage{soul}					
\usepackage{bbold}				
\usepackage{multicol}				
\usepackage{tikz-cd}
\usepackage{rotating}

\usetikzlibrary{arrows}

\tikzset{
commutative diagrams/.cd,
arrow style=tikz,
diagrams={>=latex}}

\usepackage{amsmath}
\usepackage{amssymb}
\usepackage{amsfonts}
\usepackage{mathtools}

\usepackage{feynmf}
\usepackage{marvosym}

\usepackage{import}

\newcommand*\xoverline[2][0.75]{%
    \sbox{\myboxA}{$\m@th#2$}%
    \setbox\myboxB\null
    \ht\myboxB=\ht\myboxA%
    \dp\myboxB=\dp\myboxA%
    \wd\myboxB=#1\wd\myboxA
    \sbox\myboxB{$\m@th\overline{\copy\myboxB}$}
    \setlength\mylenA{\the\wd\myboxA}
    \addtolength\mylenA{-\the\wd\myboxB}%
    \ifdim\wd\myboxB<\wd\myboxA%
       \rlap{\hskip 0.5\mylenA\usebox\myboxB}{\usebox\myboxA}%
    \else
        \hskip -0.5\mylenA\rlap{\usebox\myboxA}{\hskip 0.5\mylenA\usebox\myboxB}%
    \fi}
\makeatother

\definecolor{cobalt}{RGB}{44, 98, 120}
\definecolor{celadon}{rgb}{0.67, 0.88, 0.69}
\definecolor{dm}{cmyk}{.20, 0, .30, 0}
\definecolor{burgundy}{rgb}{0.5, 0.0, 0.13}
\definecolor{plotBlue}{RGB}{94, 130, 181}

\hypersetup{
  colorlinks,
  citecolor=Violet,
  linkcolor=cobalt,
  urlcolor=Blue}

\DeclareSymbolFontAlphabet{\mathbb}{AMSb}

\newif\iffastcompile

\fastcompilefalse

\iffastcompile
\newcommand{\mk}[1]{}
\newcommand{\lm}[1]{}
\else
\newcommand{\mk}[1]{\todo[color=burgundy!30, size=\scriptsize, bordercolor=burgundy!30]{MK: #1}}
\newcommand{\lm}[1]{\todo[color=dm!90, size=\scriptsize, bordercolor=dm!90]{LM: #1}}

\fi

\ProvideTextCommandDefault{\Dbar}{%
\leavevmode\lower.5ex\rlap{\hskip-.07em\accent"16}D%
}

\usepackage{environ}
\usepackage{changepage}

\begin{document}
	\newcommand{\main}{.}
\begin{titlepage}

\setcounter{page}{1} \baselineskip=15.5pt \thispagestyle{empty}
\setcounter{tocdepth}{2}

\bigskip\
{\hfill \small MIT-CTP/5518}
\vspace{0.5cm}
\vspace{1cm}
\begin{center}
{\LARGE \bfseries D-instanton, threshold corrections, and topological string.}
\end{center}

\vspace{0.55cm}

\begin{center}
\scalebox{0.95}[0.95]{{\fontsize{14}{30}\selectfont Manki Kim$^{a}$\vspace{0.25cm}}}

\end{center}

\begin{center}

\vspace{0.15 cm}
{\fontsize{11}{30}
\textsl{$^{a}$Center for Theoretical Physics, Department of Physics, Massachusetts Institute of Technology, Cambridge, MA 02139}}\\
\vspace{0.25cm}

\vskip .5cm
\end{center}

\vspace{0.8cm}
\noindent

In this note, we prove that the one-loop pfaffian of the non-perturbative superpotential generated by Euclidean D-branes in type II compactifications on orientifolds of Calabi-Yau threefolds is determined by the moduli integral of the new supersymmetric index defined by Cecotti, Fendley, Intriligator, and Vafa. As this quantity can be computed via topological string theory, Chern-Simons theory, matrix models, or by solving the holomorphic anomaly equation, this result provides a method to directly compute the one-loop pfaffian of the non-perturbative superpotential. The relation between the one-loop pfaffian, threshold corrections to the gauge coupling, and the one-loop partition function of open topological string theory is discussed.

\vspace{1.1cm}

\vspace{3.1cm}

\noindent\today

\end{titlepage}

\section{Introduction}
One of the most pressing questions in string theory is to construct or disprove the existence of (meta)-stable cosmological vacua with features that resemble our own universe. As an intermediate step towards understanding non-supersymmetric vacua of string theory, one can study four-dimensional $\mathcal{N}=1$ supersymmetric vacua of string theory. As the vacuum structure of such vacua is determined by the K\"{a}hler potential and the superpotential of the low-energy supergravity, it is therefore necessary to understand how to precisely compute the K\"{a}hler potential and the superpotential. 

In this note, we will study the non-perturbative terms in the superpotential generated by Euclidean D-branes in type II string theory compactified on orientifolds of Calabi-Yau threefolds \cite{Witten:1996bn}. The non-perturbative superpotential due to a Euclidean D-brane wrapped on a cycle $\Gamma$ reads
\begin{equation}
W\supset\mathcal{A}_\Gamma e^{-\mathcal{T}_\Gamma}\,,
\end{equation}
where $\mathcal{T}_\Gamma$ is the holomorphic worldvolume action of the D-instanton, and $\mathcal{A}_\Gamma$ is the one-loop pfaffian which we will also oftentimes call the one-loop prefactor. The one-loop pfaffian $\mathcal{A}_\Gamma$ can depend on moduli fields. Non-perturbative corrections to the superpotential play prominent roles in moduli stabilization scenarios \cite{Kachru:2003aw,Balasubramanian:2005zx}, and particle physics applications of string theory \cite{Abel:2006yk,Akerblom:2006hx,Blumenhagen:2006xt,Ibanez:2006da,Cvetic:2007ku}.\footnote{For review on the D-instanton effects in string compactifications, see \cite{Blumenhagen:2009qh}.} 

Therefore, it is extremely important to carefully examine and make progress on the D-instanton effects to the superpotential. To simplify the discussion, we will focus on Euclidean D-branes that only have the universal zero modes describing their positions in the non-compact directions and their superpartners. Furthermore, we will assume that the RR tadpoles are saturated by spacetime-filling D-branes to avoid the technical complications induced by the RR flux.

Despite its importance, it has been extremely challenging to compute the one-loop prefactor  $\mathcal{A}_\Gamma.$ The technical challenge is in part due to the fact that the na\"{i}ve scattering amplitudes involving the D-instantons suffer from IR divergences caused by the zero modes of the D-instantons. This may explain why most of the existing computations of the variations of the one-loop prefactor relied on indirect approaches, rather than direct computations. For example, in heterotic string theory, one can determine the functional form, but not the overall scale, of the one-loop pfaffian by carefully examining the locus in moduli space at which the number of the zero modes increases \cite{Buchbinder:2002ic,Buchbinder:2002pr}. This result was conjectured to be generalized to F-theory constructions which admit stable degenerations \cite{Cvetic:2012ts}. Similarly, in type IIB string theory and F-theory counting of the zero modes from open strings extended between the D-instanton and spacetime filling D-brane led to the computation of the one-loop prefactor \cite{Ganor:1996pe,Baumann:2006th,Kim:2022uni}, but again the overall scale was left undetermined. The result of \cite{Ganor:1996pe} is confirmed through hard computations in toroidal orientifold compactifications \cite{Berg:2004ek}.\footnote{For analogous computations, see \cite{Billo:2007py}.} If the non-perturbative superpotential in question is generated by the D(-1)-instantons, then algebro-geometric techniques can be used to determine the D(-1)-instanton corrections without ambiguities \cite{Kim:2022jvv}.\footnote{The superpotential due to D3-D(-1) bound states at orbifold singularities was studied in \cite{Billo:2008pg}.} From the M-theoretic approach, one can study the intermediate Jacobian of the Euclidean M5-branes to constrain the one-loop pfaffian \cite{Witten:1996hc}.\footnote{For the computation of the intermediate Jacobian using algebro geometric techniques, see for example \cite{Denef:2005mm,Blumenhagen:2010ja,Kim:2021lwo,Jefferson:2022ssj}.} But this approach also lacks the capability to determine the overall scale, and at the same time, it is very difficult to explicitly compute the intermediate Jacobian explicitly except for very special cases \cite{Grimm:2011dj,Kerstan:2012cy}. 

Recently, in a series of works on D-instanton amplitudes from the point of view of string field theory \cite{Balthazar:2019rnh,Sen:2019qqg,Balthazar:2019ypi,Sen:2021qdk,Sen:2021tpp,Sen:2021jbr,Alexandrov:2021shf,Alexandrov:2021dyl,Agmon:2022vdj}, it was realized that string field theory can be used to regulate the IR divergence of the D-instanton amplitudes. In particular, the IR-regulated D-instanton amplitudes were found to be in perfect agreement with the predictions from mirror symmetry, S-duality, and twistorial description of quaternionic geometries \cite{Alexandrov:2021shf,Alexandrov:2021dyl}. Prompted by this success, in \cite{Alexandrov:2022mmy} the D-instanton correction to the superpotential was computed in terms of the open string spectrum. Because of the lack of the predictive power of dualities and supersymmetry in 4d $\mathcal{N}=1$ theories, the one-loop prefactor computed in \cite{Alexandrov:2022mmy} has not been crosschecked yet. But, as \cite{Alexandrov:2022mmy} followed the same prescription as \cite{Alexandrov:2021shf,Alexandrov:2021dyl}, it is strongly suggestive that the overall normalization obtained by \cite{Alexandrov:2022mmy} is correct. It is important to perform an independent crosscheck of the results of \cite{Alexandrov:2022mmy}.

But, the results of \cite{Alexandrov:2022mmy} may not be fully satisfactory because, in Calabi-Yau compactifications, the spectrum of strings is not always accessible. In fact, in the case of Calabi-Yau compactifications, it is not even clear how to approximate the string spectrum away from the large volume limit. Therefore, one can reasonably complain that \cite{Alexandrov:2022mmy} has replaced a practically-impossible-to-compute-quantity with another practically-impossible-to-compute-quantity.

In this note, we will show that in fact the one-loop prefactor of the non-perturbative superpotential is more computable than one na\"{i}vely would have thought even if the access to the spectrum of the Calabi-Yau CFT is limited. 

A crucial insight comes from the holomorphic anomaly equations and topological string amplitudes studied by Bershadsky, Cecotti, Ooguri, and Vafa (BCOV) in a series of papers \cite{Bershadsky:1993ta,Bershadsky:1993cx}. In \cite{Bershadsky:1993cx}, it was conjectured that the topological string partition function computes F-terms in the low-energy supergravity theories derived from string compactifications. In the closed string case, this conjecture was proved by direct computations \cite{Antoniadis:1993ze} that indeed the topological string partition function at genus g computes the following F-term
\begin{equation}
\int d^4\theta F_g (\mathcal{W}^2)^{g}\,,
\end{equation}
where $\mathcal{W}$ is the square of the Weyl superfield of $\mathcal{N}=2$ supergravity multiplet. The open string version of the conjecture that the open string topological partition function with h holes computes the F-term of the form
\begin{equation}
\int d^2\theta F_{0,h} (\text{Tr}(W)^2)^{h-1}\,,
\end{equation}
where $W$ is the chiral superfield for the gauge field strength, was confirmed in type I string theory \cite{Antoniadis:2005sd}. In particular, at the one-loop level $h=2,$ the open topological partition function is conjectured to compute the threshold correction to the gauge coupling\footnote{In the context of $E_8\times E_8$ heterotic string compactifications with the standard embedding \cite{Candelas:1985en}, it was proven in \cite{Antoniadis:1992rq,Antoniadis:1992sa,Kaplunovsky:1995jw} that the closed topological partition function computes the difference between the gauge threshold corrections to $E_6$ and $E_8$ gauge groups.} and the one-loop partition function is written as 
\begin{equation}
F_{0,2}=\int\frac{dt}{t} \text{Tr}_R\left( (-1)^F F q^{L_0-\frac{3}{8}}\right)\,, 
\end{equation}
which is the moduli integral of the new supersymmetric index defined by \cite{Cecotti:1992qh}, which we will call the CFIV index. Although it is conceivable that the open string version of the BCOV conjecture should hold in all type II compactifications on orientifolds of Calabi-Yau manifolds, this is not yet confirmed.

The other thread of insight comes from the relationship between the threshold correction to the gauge coupling on a spacetime filling D-brane and the one-loop prefactor $\mathcal{A}_\Gamma$ of the D-instanton superpotential \cite{Abel:2006yk,Akerblom:2006hx}. Because the D-instanton can be understood as a gauge instanton on a spacetime filling D-brane \cite{Atiyah:1978ri,Witten:1994tz,Douglas:1995bn,Douglas:1996uz}, it is natural to conjecture that the exponentiated threshold correction is equivalent to the one-loop prefactor of the D-instanton superpotential. And in fact, in the type II compactifications on orientifolds of Calabi-Yau threefolds, it was proven in \cite{Alexandrov:2022mmy} that the exponentiated threshold correction to a probe spacetime filling D-brane wrapped on a cycle $\Gamma$ is exactly the same as the one-loop prefactor $\mathcal{A}_\Gamma.$\footnote{In \cite{Akerblom:2006hx,Abel:2006yk}, this statement was shown for the annuli contributions, but not for the M\"{o}bius strip contribution.}

Continuing this train of thought, it is then only natural to conjecture that the one-loop pfaffian of the non-perturbative superpotential is determined by the CFIV index\footnote{In the context of $\mathcal{N}=2$ supersymmetric gauge theories, it was observed that an $\mathcal{N}=2$ index captures BPS instantons \cite{Alexandrov:2014wca}. It would be interesting to see a connection between topologial string partition function and the phenomena observed in \cite{Alexandrov:2014wca}. We thank Sergei Alexandrov for pointing out this interesting result.}
\begin{equation}
\mathcal{A}_\Gamma \simeq e^{F_{0,2}}\,.
\end{equation}
Note that the above equation cannot be quite correct, as the topological string partition function suffers from holomorphic anomaly. We will make the correspondence between $\mathcal{A}_\Gamma$ and $F_{0,2}$ more precise later in this paper. This possibility was already contemplated in \cite{Walcher:2007qp}. In this note, with the help of the results of \cite{Alexandrov:2022mmy} and the character formulas of the extended $\mathcal{N}=2$ superconformal algebra studied in \cite{Odake:1988bh,Odake:1989dm,Odake:1989ev}, we will prove that the one-loop prefactor of the nonperturbative superpotential is determined by the new supersymmetric index. This also proves the BCOV conjecture at the one-loop level for D-brane gauge theories with no matter. As the moduli integral of the CFIV index is computable via open topological string theory and holomorphic anomaly equations \cite{Walcher:2006rs,Walcher:2007tp,Morrison:2007bm,Walcher:2007qp,Bonelli:2009aw,Bonelli:2010cu}, this result paves a way to direct computation of the one-loop pfaffian $\mathcal{A}_\Gamma.$

This note is organized as follows. In \S\ref{sec:D-instanton}, we prove that the one-loop pfaffian is determined by the CFIV index. 
In \S\ref{sec:conclusion} we conclude. In appendices, we collect useful formulas. In \S\ref{app:theta}, we collect useful formulas involving the Jacobi theta functions and the Dedekind eta function. In \S\ref{app:character}, we collect the character formulas of the extended $\mathcal{N}=2$ superconformal algebra.
\section{D-instanton superpotential and the CFIV index.}\label{sec:D-instanton}
We shall study type II string theory compactified on a Calabi-Yau threefold $X.$ The worldsheet CFT is decomposed into the $b, c, \beta, \gamma$ ghost CFT, the free field CFT with central charge $(c,\bar{c})=(6,6)$ and $(1,1)$ supersymmetry describing the four non-compact directions, and strongly coupled CFT with central charge $(c,\bar{c})=(9,9)$ and $(\mathcal{N},\bar{\mathcal{N}})=(2,2) $ supersymmetry describing the Calabi-Yau non-linear sigma model. We will oftentimes denote the strongly coupled Calabi-Yau CFT by \emph{internal} CFT. We will consider an orientifolding of this worldsheet CFT. The details of the orientifolding can be found in \cite{Alexandrov:2022mmy}.

We study a Euclidean D-brane wrapped on a cycle $\Gamma.$ We will assume that the Euclidean D-brane only has the universal zero modes and none other to ensure that the non-perturbative superpotential is generated. In \cite{Alexandrov:2022mmy}, by carefully studying the D-instanton scattering amplitudes and its relation to the one-loop partition function of open string field theory, the non-perturbative superpotential generated by the Euclidean D-brane was determined\footnote{For earlier work, see \cite{Blumenhagen:2006xt}.}
\begin{equation}
|W_{\Gamma}|= \frac{\kappa_4^3}{16\pi^2} e^{-\mathcal{K}/2} K_0\text{Re}(\mathcal{T}_\Gamma) |e^{-\mathcal{T}_\Gamma}|\,,
\end{equation}
where $\mathcal{T}_{\Gamma}$ is the disk level effective action of the D-instanton, $\mathcal{K}$ is the tree-level K\"{a}hler potential, and we define
\begin{equation}\label{eqn: K0}
K_0:=\lim_{\epsilon\rightarrow 0}\lim_{\epsilon'\rightarrow 0}\exp\left[\int_{\epsilon'}^{1/\epsilon}\frac{dt}{2t} Z_A+\int_{\epsilon'/4}^{1/\epsilon}\frac{dt}{2t} Z_M+3\int_0^{1/\epsilon}\frac{dt}{2t} (e^{-2\pi t}-1)\right]\,.
\end{equation}
In \eqref{eqn: K0}, $Z_A$ is the sum of annuli diagrams with at least one end on the D-instanton, and $Z_M$ is the M\"{o}bius strip diagram with the end on the D-instanton. The extra factor $\frac{1}{2}$ was included in $Z_A$ and $Z_M$ due to the orientifold projection. The last term acts as the IR regulator for the IR divergence that arises from the zero modes of the D-instanton. Therefore, to compute the non-perturbative superpotential precisely, we must compute the annuli diagrams and the M\"{o}bius strip diagram. This is the focus of this section.

Let us now study the one-loop diagrams with one end on the D-instanton and the other end on a spacetime filling D-brane.\footnote{The contribution from the annulus diagram with both ends on the D-instanton vanishes due to higher supersymmetry. This was also proven in \cite{Alexandrov:2022mmy}. Therefore, we will not explicitly consider this diagram in this note.} In \cite{Alexandrov:2022mmy}, the contribution from an orbit generated by the integral spectral flow whose highest weight state is a massive state with $(h,Q)$ of $\mathcal{N}=2$ superconformal algebra was determined to be
\begin{equation}\label{eqn:DN threshold}
Z_{A}^{(h,Q)}= \frac{(-1)^Q}{2} q^{h-\frac{1+Q}{4}}\vartheta_{1-Q,0}(2\tau)\,.
\end{equation}
Note here that $h$ is eigenvalue of $L_0$ of the Virasoro algebra, and $Q$ is eingenvalue of $I_0,$ the $U(1)_R$ charge of the extended superconformal algebra. For the details of the extended superconformal algebra see \S\ref{app:character}. We can rewrite \eqref{eqn:DN threshold} as
\begin{equation}\label{eqn:DN threshold2}
(-1)^{Q+1} \frac{1}{4\pi \eta(\tau)^3} q^{h-\frac{1+Q}{4}} \left(-2\pi \eta(\tau)^3\right) \vartheta_{1-Q,0}(2\tau)\,.
\end{equation}
By using an identity
\begin{equation}
\partial_z \vartheta_{11}(z|\tau)|_{z=0}=-2\pi \eta(\tau)^3\,,
\end{equation}
we conclude that the following identity holds
\begin{align}
Z_{A}^{(h,Q)}=\frac{(-1)^{Q+1}}{4\pi \eta(\tau)^3} q^{h-\frac{1+Q}{4}}\vartheta_{1-Q,0}(2\tau) \partial_z \vartheta_{1,1}(z|\tau)|_{z=0} \,.
\end{align}
We use the following identity, c.f. \eqref{eqn:dev1}-\eqref{eqn:dev2},
\begin{equation}
\frac{1}{2\pi i} \partial_z ch_{\tilde{R}}^{(h,Q)}(z,\tau)|_{z=0}=\frac{(-1)^{Q+1}}{2\pi \eta(\tau)^3} q^{h-\frac{1+Q}{4}}\vartheta_{1-Q,0}(2\tau) \partial_z \vartheta_{1,1}(z|\tau)|_{z=0}\,,
\end{equation}
to rewrite
\begin{equation}
Z_{A}^{(h,Q)}=\frac{1}{2} \left[\frac{1}{2\pi i}\partial_z ch_{\tilde{R}}^{(h,Q)}(z,\tau)|_{z=0}\right]\,.
\end{equation}
Because we assumed that the only zero modes of the D-instanton are the universal zero modes, massless representations of the annuli diagrams are absent. 

For next step, we shall relate $I_0$ with the fermion number operator. The central idea behind this identification is that the $U(1)_R$ charge of the states in the Calabi-Yau CFT was used to perform the GSO projection. We will, therefore, identify the following generators  \cite{Lerche:1989uy,Odake:1989ev}
\begin{equation}
F\equiv I_0\,.
\end{equation}
This identification is well justified for the following reason. In $\mathcal{N}=2$ superconformal theories, the Witten index
\begin{equation}
\text{Tr}_R\left((-1)^{F-\frac{3}{2}}\right)
\end{equation}
is computed by
\begin{equation}
\text{Tr}_R\left((-1)^{I_0-\frac{3}{2}}\right)\,,
\end{equation}
as was studied in \cite{Odake:1988bh}. This justifies the identification $(-1)^F\equiv (-1)^{I_0}.$ Let us look at one more evidence. We note that the commutators between the fermion number operator and the supercurrents $G$ and $\tilde{G}$ are, c.f. \eqref{eqn:charge1} and \eqref{eqn:charge2},
\begin{equation}
[F,G]=G\,,\quad [F,\tilde{G}]=-\tilde{G}\,.
\end{equation}
These equations again justfy the identification
\begin{equation}
I_0\equiv F \mod 2\,.
\end{equation}

But, the arguments we have presented so far do not justify the identification $I_0\equiv F$ yet. To fully fix the identification between $I_0$ and the fermion number, we note that the vacuum state in the NS-sector that corresponds to an identity operator has $Q=0.$ Because this state has zero worldsheet fermion number, we conclude $I_0\equiv F.$ Note that this identification was also used in \cite{Antoniadis:1992rq,Antoniadis:1992sa,Kaplunovsky:1995jw} to reproduce the one loop partition function of closed topological string theory. In the NS sector, we can therefore identify
\begin{equation}
I_0\equiv F\,,
\end{equation}
for all states. Note that in the R-sector, the definition of the fermion number is more subtle. To avoid this subtlety, in this note, we will use the spectral flow whenever necessary to determine the fermion number in the R-sector. 

Now note that, as we reviewed in \S\ref{app:character}, the character formula in the $\tilde{R}$ sector is given as
\begin{equation}
ch^{(h,Q)}_{\tilde{R}}(z,\tau)= \text{Tr}_R\left((-1)^{F-\frac{3}{2}} q^{L_0-\frac{3}{8}} e^{2\pi i z F}\right)\,,
\end{equation}
see \eqref{eqn:spectral 3}. Therefore, taking a derivative in $z$ simply brings down a factor of $2\pi i F$ in the trace
\begin{equation}
\frac{1}{2\pi i}\frac{\partial}{\partial z}ch^{(h,Q)}_{\tilde{R}}(z,\tau)= \text{Tr}_R\left((-1)^{F-\frac{3}{2}}F q^{L_0-\frac{3}{8}} e^{2\pi i z F}\right)\,.
\end{equation}

As the sum of $\frac{1}{2\pi i} \partial_z ch_{\tilde{R}}^{(h,Q)}(z,\tau)|_{z=0}$ is by definition the new supersymmetric index defined by Cecotti, Fendley, Intriligator, and Vafa \cite{Cecotti:1992qh}
\begin{equation}
\text{Tr}_R \left( (-1)^{F-\frac{3}{2}} F q^{L_0-\frac{3}{8}}\right)= \sum_{(h,Q)}\frac{1}{2\pi i} \partial_z ch_{\tilde{R}}^{(h,Q)}(z,\tau)|_{z=0}\,
\end{equation}
we arrive at one of our main equations 
\begin{equation}\label{eqn:top1}
Z_{A}=\sum_{(h,Q)}Z_{A}^{(h,Q)}=\frac{1}{2} \text{Tr}_R \left( (-1)^{F-\frac{3}{2}} F q^{L_0-\frac{3}{8}}\right)\,,
\end{equation}
where the trace is taken over the states of the internal CFT.\footnote{We thank Sergei Alexandrov for catching the typo.} As a result, we obtained that the one-loop partition function is precisely half of the new supersymmetric index! An important remark should follow. As we reviewed  in \S\ref{app:character}, we obtained the states in the Ramond sector by applying the half-integral spectral flow to states in the NS sector labeled by $(h,Q).$ Although the eigenvalues of $(L_0,J)$ change under the spectral flow, for the notational simplicity, we are keeping track of $(h,Q)$ of the original state in the NS sector.

We now prove a similar statement for the M\"{o}bius strip diagram. As was proven in \cite{Alexandrov:2022mmy}, the contribution from a massive representation with $(h,Q)$ to the M\"{o}bius diagram is given by
\begin{equation}\label{eqn:mobius}
Z_M^{(h,Q)}=(-1)^{1-Q}\alpha_{(h,Q)} \hat{q}^{h-\frac{1+Q}{4}}\vartheta_{1-Q,0}(2\hat{\tau})\,,
\end{equation}
where $\alpha_{(h,Q)}$ is a phase induced by the orientifold action, $\hat{\tau}=\tau+1/2,$ and $\hat{q}=\exp(2\pi i\hat{\tau}).$ We now again see that \eqref{eqn:mobius} is the same as
\begin{equation}\label{eqn:mobius2}
-\frac{\alpha_{(h,Q)}}{2\pi i} \partial_zch_{\tilde{R}}^{(h,Q)}(z,\hat{\tau})|_{z=0}\,.
\end{equation}
Note again that the $(h,Q)$ should be understood as $(h,Q)$ of the highest weight in the NS-sector. Unlike in the case of annuli diagrams, we have included an extra negative sign because the orientifold action flips the sign of the vacuum. As we assumed that the only zero modes of the D-instanton are the universal zero modes, we only need to consider the vacuum representation among the massless representations. Because the character of the vacuum representation can be written in terms of the character formulas of massive representations, c.f. \eqref{eqn:massless characters}, we again reach the same conclusion. Thus, we conclude that the following equation holds
\begin{equation}\label{eqn:mobius3}
Z_M=\sum_{(h,Q)}Z_M^{(h,Q)}=  \text{Tr}_R\left( (-1)^{F-\frac{3}{2}} F \Omega q^{L_0-\frac{3}{8}}\right)\,,
\end{equation}
where the trace is taken over the states of the internal CFT, and $\Omega$ is the orientifold projection.\footnote{It is interesting to observe that there is an extra factor of $1/2$ in \eqref{eqn:top1} compared to \eqref{eqn:mobius3}. This is due to the fact that a single spacetime D-brane in the Calabi-Yau should be understood as a fractional D-brane in the orientifold. For example, a seven-brane stack with SO(8) gauge group can be understood as a bound-state of eight seven branes and an O7-plane in a Calabi-Yau. But, the same configuration should be understood as a bound-state of four seven-branes and an O7-plane in the orientifold.} Note that we used the defintion
\begin{equation}\label{eqn:id1}
\text{Tr}_R \left( (-1)^{F-\frac{3}{2}} F\Omega q^{L_0-\frac{3}{8}}\right)=- \sum_{(h,Q)}\frac{\alpha_{(h,Q)}}{2\pi i} \partial_z ch_{\tilde{R}}^{(h,Q)}(z,\tau)|_{z=0}\,,
\end{equation}
which also serves as a definition for the phase induced by the orientifold action $\Omega$ on states in the R-sector.

Now let us perform a consistency check of the formula \eqref{eqn:mobius3}. The most basic check is to confirm the consistency of the zero modes contributions to \eqref{eqn:mobius3}. As we assumed that the only zero modes of the D-instanton are four bosonic and two fermionic zero modes, we expect to see $3$ in $Z_M$ in $t\rightarrow \infty $ limit. Let us check this conclusion from equations \eqref{eqn:mobius} and \eqref{eqn:mobius2}. As was studied in \cite{Alexandrov:2022mmy}, $\alpha$ for the vacuum representation was determined to be $-1,$ and therefore twice the contribution from the vacuum representation is given by
\begin{equation}
\hat{q}^{-\frac{1}{4}} \vartheta_{1,0}(2\hat{\tau})+\vartheta_{0,0}(2\hat{\tau})\,.
\end{equation}
As in $\hat{q}\rightarrow 0$ limit, $\vartheta_{1,0}(2\hat{\tau})=2\hat{q}^{1/4}$ and $\vartheta_{0,0}(2\hat{\tau})=1,$ we correctly reproduce 3 in $t\rightarrow \infty$ limit.

Now we shall attempt to reproduce this result by a manual computation in the R-sector using \eqref{eqn:mobius3}. There are two massless states in the R-sector, and both of which belong to the orbit connected to the vacuum representation by the half-integral spectral flow. The massless states are the one with $(h,Q)=(3/8,3/2),$ and the other with $(h,Q)=(3/8,-3/2).$ Let $X^R$ and $\tilde{X}^R$ be the integral spectral flow operators in the R-sector. We choose a convention such that $X^R$ is connected to $X$ with $(h,Q)=(3/2,3),$ and $\bar{X}^R$ is connected to $\tilde{X}$ with $(h,Q)=(3/2,-3).$ Then, we have \cite{Odake:1988bh,Odake:1989dm}
\begin{equation}
\frac{1}{\sqrt{2}}X_0^R\left|\frac{3}{8},-\frac{3}{2}\right\rangle=\left|\frac{3}{8},\frac{3}{2}\right\rangle\,,
\end{equation}
and
\begin{equation}
\frac{1}{\sqrt{2}}\tilde{X}_0^R\left|\frac{3}{8},\frac{3}{2}\right\rangle=\left|\frac{3}{8},-\frac{3}{2}\right\rangle\,.
\end{equation}
As was studied in \cite{Alexandrov:2022mmy}, action of the orientifolding can be fixed so that the resulting theory in the target space preserves $\mathcal{N}=1$ supersymmetry. The result is that the orientifolding commutes with $X_0$ and $\tilde{X}_0$ \cite{Alexandrov:2022mmy}, we conclude that the phases generated by the orientifold action are the same between the states $(3/8,3/2)$ and $(3/8,-3/2).$ This leads to a manual computation of the contribution of the massless state
\begin{equation}
\text{Tr}_R\left( (-1)^{F-\frac{3}{2}} F\Omega q^{L_0-\frac{3}{8}}\right)= ( 3/2-(-3/2))=3\,,
\end{equation}
therefore reproducing 3.\footnote{We thank Ashoke Sen and Sergei Alexandrov for illuminating discussions.}

\section{Conclusions}\label{sec:conclusion}
In this note, we proved that the one-loop prefactor $\mathcal{A}_\Gamma$ of the non-perturbative superpotential is determined by the CFIV index. Because the moduli integral of the CFIV index can be computed via various techniques including topological string theory, holomorphic anomaly equations \cite{Walcher:2007tp,Walcher:2007qp}, Chern-Simons theory \cite{Ooguri:1999bv}, and matrix models \cite{Marino:2006hs,Eynard:2007hf}, the results of this note provide a principled way to evaluate the one-loop prefactor in generic Calabi-Yau compactifications.

There are a few interesting future directions one can pursue. 
\begin{itemize}
\item The most imminent next step is to compute the one-loop prefactor of the D-instanton superpotential in explicit examples. One imminent technical challenge is to understand the universal behavior of the holomorphic ambiguities around singular points in the moduli space for annuli and the M\"{o}bius strip diagrams. 
\item In \cite{Witten:1996hc}, Witten showed that if a Euclidean M5-brane wrapped on a divisor in a Calabi-Yau fourfold has a trivial intermediate Jacobian, then the one-loop prefactor does not depend on moduli in the \emph{absence} of spacetime filling M2-branes. In type IIB compactification on O3/O7-orientifolds, this implies that Euclidean D3-branes and gaugino condensations which are dual to such Euclidean M5-branes generate the non-perturbative superpotential terms that do not depend on moduli in the absence of D3-brane contributions. As this statement is inherently topological, it may be possible to confirm it by using this paper's result. 
\item As the one-loop prefactor can be computed in heterotic string compactifications \cite{Buchbinder:2002ic,Buchbinder:2002pr} and its extension to F-theory compactifications \cite{Cvetic:2012ts}, it would be interesting to crosscheck the results via heterotic/type II dualities.
\end{itemize}
\section*{Acknowledgements}
The work of MK was supported by the Pappalardo fellowship. We thank Ashoke Sen and Sergei Alexandrov for their valuable comments. We thank Sergei Alexandrov for catching typos in the manuscript. We thank Atakan Hilmi Fırat, Liam McAllister, Jakob Moritz, and Andreas Schachner for discussions.
\appendix
\newpage
\section{The Jacobi theta functions}\label{app:theta}
In this section, we collect useful formulas for the Jacobi theta functions and the character formulas for $\mathcal{N}=2$ superconformal algebra. In this paper, we will mostly follow the conventions of \cite{Alexandrov:2022mmy}. We define
\begin{equation}
\vartheta_{\alpha,\beta}(z|\tau)= \sum_{n\in \Bbb{Z}+\frac{\alpha}{2}} e^{i\pi n \beta} q^{n^2/2}y^n\,,\quad q=e^{2\pi i\tau}\,,\quad y=e^{2\pi iz}\,.
\end{equation}
For $\vartheta_{\alpha,\beta}(0|\tau),$ we use a shorthand notation $\vartheta_{\alpha,\beta}(\tau):=\vartheta_{\alpha,\beta}(0|\tau).$ We collect $\vartheta_{\alpha,\beta}$ for $(\alpha,\beta)=(0,0),(0,1),(1,0),(1,1)$ 
\begin{align}
\vartheta_{0,0}(z|\tau)=&\prod_{n=1}^{\infty}(1-q^n) \left(1+(y+y^{-1})q^{n-\frac{1}{2}}+q^{2n-1}\right)\,,\\
\vartheta_{0,1}(z|\tau)=&\prod_{n=1}^{\infty}(1-q^n)\left(1-(y+y^{-1})q^{n-\frac{1}{2}}+q^{2n-1}\right)\,,\\
\vartheta_{1,0}(z|\tau)=&q^{\frac{1}{8}}(y^{\frac{1}{2}}+y^{-\frac{1}{2}})\prod_{n=1}^\infty (1-q^n)\left(1+(y+y^{-1})q^n+q^{2n}\right)\,,\\
\vartheta_{1,1}(z|\tau)=&iq^{\frac{1}{8}}(y^{\frac{1}{2}}-y^{-\frac{1}{2}})\prod_{n=1}^\infty (1-q^n)\left(1-(y+y^{-1})q^n+q^{2n}\right)\,.
\end{align}
The Jacobi theta functions admit quasi-periodicity, 
\begin{align}
\vartheta_{\alpha,\beta}\left(z+\frac{1}{2}|\tau\right)=&\vartheta_{\alpha,\beta+1}(z|\tau)\,,\\
\vartheta_{\alpha,\beta}\left(z+\frac{\tau}{2}|\tau\right)=&e^{-i\pi \beta/2}q^{-\frac{1}{8}}y^{-\frac{1}{2}}\vartheta_{\alpha+1,\beta}(z|\tau)\,,\\
\vartheta_{\alpha+2,\beta}(z|\tau)=&\vartheta_{\alpha,\beta}(z|\tau)\,,\\
\vartheta_{\alpha,\beta+2}(z|\tau)=&e^{i\alpha\pi}\vartheta_{\alpha,\beta}(z|\tau)\,.
\end{align}
Theta functions satisfy so-called the Jacobi identity
\begin{equation}
\vartheta_{0,0}(\tau)^4-\vartheta_{0,1}(\tau)^4-\vartheta_{1,0}(\tau)^4=0\,.
\end{equation}

We also record useful relations between the Jacobi theta functions and the Dedekind eta function. We define the Dedekind eta function as
\begin{equation}
\eta(\tau)=q^{\frac{1}{24}}\prod_{n=1}^\infty (1-q^n)\,.
\end{equation}
The Dedekind eta function satisfies the following identities
\begin{equation}
\partial_z \vartheta_{1,1}(z|\tau)|_{z=0}=-2\pi \eta(\tau)^3\,,
\end{equation}
and
\begin{align}
\vartheta_{1,0}(\tau)=&\frac{2\eta(2\tau)^2}{\eta(\tau)}\,,\\
\vartheta_{0,1}(\tau)=&\frac{\eta(\frac{1}{2}\tau)^2}{\eta(\tau)}\,,\\
\vartheta_{0,0}(\tau)=&\frac{\eta(\tau)^5}{\eta(\frac{1}{2}\tau)^2\eta(2\tau)^2}\,.
\end{align}
\section{Character formulas for $\mathcal{N}=2$ superconformal algebra}\label{app:character}

In this section, we will summarize important properties of $\mathcal{N}=2$ superconformal algebra and its associated representation theory \cite{Eguchi:1988vra,Odake:1988bh,Odake:1989ev,Odake:1989dm}. In this section we will focus on the left moving sector. The superconformal algebra of the right moving sector can be obtained by taking the complex conjugate of the algebra of the left moving sector. As we are interested in Calabi-Yau threefold compactifications, we will restrict to $c=9.$

We first collect OPEs for superconformal generators, the energy-momentum tensor $T,$ super-currents $G$ and $\tilde{G},$ and U(1) current $I,$ 
\begin{align}
T(z)T(w)=&\frac{c}{2(z-w)^4}+\frac{2T(w)}{(z-w)^2}+\frac{\partial T(w)}{z-w}+\dots\,,\\
I(z)I(w)=&\frac{c}{3(z-w)^2}+\dots\,,\\
I(z)G(w)=&\frac{1}{z-w}G(w)+\dots\,,\label{eqn:charge1}\\
I(z)\tilde{G}(w)=&-\frac{1}{z-w}\tilde{G}(w)+\dots\,,\label{eqn:charge2}\\
G(z)\tilde{G}(w)=&\frac{2c}{3(z-w)^3}+\frac{2I(w)}{(z-w)^2}+\frac{1}{z-w}(\partial I(w)+2T(w))+\dots\,,\\
G(z)G(w)=&\text{regular}\,,\\
\tilde{G}(z)\tilde{G}(w)=&\text{regular}\,.
\end{align}
As was studied in \cite{Odake:1988bh}, it was shown that $\mathcal{N}=2$ superconformal algebra extends to the extended superconformal algebra by including the spectral flow generators $X,\tilde{X}$ and their superpartners $Y, \tilde{Y}.$\footnote{$X$ and $\tilde{X}$ generate integral shifts of the spectral flow.}

We now summarize the character formulas for the extended superconformal algebra. The character is defined as the partition function of the orbit, generated by integral spectral flow, of an irreducible representation of the extended superconformal algebra. For an irreducible representation $r,$ we define the character to be
\begin{equation}
ch_{\bullet}^{(r)}(z,\tau):=\text{Tr}_{\bullet,r} \left(q^{L_0-\frac{3}{8}} y^{I_0}\right)\,,
\end{equation}
where $\bullet$ can be either Neveu-Schwarz (NS) sector or Ramond (R) sector. Note that we defined 
\begin{equation}
q:= e^{2\pi i\tau}\,,
\end{equation}
and
\begin{equation}
y:=e^{2\pi i z}\,.
\end{equation}
The character with the GSO projection will be denoted as
\begin{equation}
ch_{\tilde{\bullet}}^{(r)}(z,\tau)\,.
\end{equation}
Note that the states in the R-sectors have half-integral charges. To take this into account, we shall include the phase shift $(-1)^{-3/2}$ in $ch_{\tilde{R}}.$ This shift agrees with the Witten index computed in \cite{Odake:1989ev}. The character formula satisfies following relations
\begin{align}
ch_{\tilde{NS}}^{(r)}(z,\tau)=&ch_{NS}^{(r)}(z+1/2,\tau)\,,\\
ch_{R}^{(r)}(z,\tau)=&q^{\frac{3}{8}}y^{\frac{3}{2}} ch_{NS}^{(r)} (z+\tau/2,\tau)\,,\\
ch_{\tilde{R}}^{(r)}(z,\tau)=& e^{-3i\pi/2} ch_{R}^{(r)}(z+1/2,\tau)\,.\label{eqn:spectral 3}
\end{align}
We will use a collection of shorthand notations
\begin{align}\label{eqn:characters}
&g_{00}^{(r)}:=\text{Tr}_{NS,r}\left( q^{L_0-\frac{3}{8}} \right)\,,\quad g_{01}^{(r)}:=\text{Tr}_{NS,r}\left( q^{L_0-\frac{3}{8}} (-1)^{I_0}\right)\,,\\
&g_{10}^{(r)}:=\text{Tr}_{R,r}\left( q^{L_0-\frac{3}{8}} \right)\,,\quad \,\,\,g_{11}^{(r)}:=\text{Tr}_{R,r}\left( q^{L_0-\frac{3}{8}} (-1)^{I_0-\frac{3}{2}}\right)\,.\label{eqn:characters2}
\end{align}
Note the shift $3/2$ in the exponent of $(-1)$ in $g_{11}^{(r)}.$ This shift is due to the definition of $ch_{\tilde{R}}^{(r)}(z,\tau).$ 

We will first study massive representations. Because representations in the R sector can be brought to representations in the NS sector by half-integral spectral flows, we will focus on irreducible representations in the NS sector. In the NS sector, the highest weight state of any massive state is constrained to have $Q=-1,0,1,$ where $h>|Q|/2.$ Because the character formula for $-Q$ is the same as the character formula for $Q,$ we will only consider $Q\geq0$ for simplicity. As a massive state is labeled by $(h,Q),$ we will denote a massive state by $(h,Q).$ The character formula for all sectors takes the form\footnote{Note that the character formulas are well defined even when $h=|Q|/2.$ This will allow us to relate the character formulas of the massles states to character formulas of the massive states in the limit $h\rightarrow |Q|/2.$}
\begin{equation}\label{eqn:character massive}
g_{\alpha\beta}^{(h,Q)}:=q^{\frac{3\alpha}{8}}g\left(\frac{\alpha\tau+\beta}{2},\tau;h,Q\right)\,,
\end{equation}
where we define
\begin{equation}
g(z,\tau;h,Q):=\frac{q^{h-\frac{1+Q^2}{4}}}{\eta(\tau)} f_{1,0}(z,\tau)f_{2,Q}(z,\tau)\,,
\end{equation}
\begin{equation}
f_{k,Q}(z,\tau):= \frac{1}{\eta(\tau)}q^{\frac{Q^2}{2k}} y^Q \vartheta_{0,0}(kz+Q\tau|k\tau)\,.
\end{equation}
\eqref{eqn:character massive} admits a simpler form\footnote{Note that the formula (D.7) in \cite{Alexandrov:2022mmy} is missing $e^{-i\pi\alpha\beta/2}.$ This error does not change the conclusions of \cite{Alexandrov:2022mmy}.}
\begin{equation}\label{eqn:character massive2}
g_{\alpha\beta}^{(h,Q)}=e^{-i\pi\alpha\beta/2} (-1)^{\beta Q} \frac{q^{h-\frac{1+Q}{4}}}{\eta(\tau)^3}\vartheta_{\alpha,\beta}(\tau)\vartheta_{\alpha+Q,0}(2\tau)\,.
\end{equation}

We shall now compute
\begin{equation}\label{eqn:dev1}
\frac{1}{2\pi i}\partial_z ch_{\tilde{R}}^{(h,Q)}(z,\tau)\,.
\end{equation}
One can write the character in the $\tilde{R}$ sector as
\begin{equation}
ch_{\tilde{R}}^{(h,Q)}(z,\tau)=q^{\frac{3}{8}} g\left(\frac{1+\tau}{2}+z,\tau;h,Q\right)\,.
\end{equation}
Because $f_{1,0}((1+\tau)/2+z,\tau)$ has a simple zero at $z=0$ due to $\vartheta_{0,0}((1+\tau)/2,\tau)=0,$ whereas $f_{2,Q}((1+\tau)/2,\tau)\neq0,$ the derivative of $ch_{\tilde{R}}^{(h,Q)}(z,\tau)$ greatly simplifies at $z=0.$ Note that if a function $G(z)$ can be written as $G(z)=A(z)B(z),$ where $A(z)$ has a simple zero at $z=0,$ and $B(z)\neq0,$ then $G'(0)=A'(0)B(0).$ Using this, we can write
\begin{equation}
\frac{\partial}{\partial z} ch_{\tilde{R}}^{(h,Q)}(z,\tau)|_{z=0}=\frac{q^{h-\frac{1+Q^2}{4}+\frac{3}{8}}}{\eta(\tau)^3}\vartheta_{0,0}'\left(\frac{1+\tau}{2}|\tau\right)(-1)q^{\frac{Q^2}{4}+\frac{Q}{2}}\vartheta_{0,0}\left(1+\tau+Q\tau|2\tau\right)\,.
\end{equation}
By using the quasi-periodicity of the theta functions, we conclude
\begin{equation}
\vartheta_{0,0}'\left(\frac{1+\tau}{2}|\tau\right)=(-i)q^{-\frac{1}{8}}\vartheta_{1,1}'(\tau)\,,
\end{equation}
and
\begin{equation}
\vartheta_{0,0}\left(1+\tau+Q\tau|2\tau\right)=(-1)^{1+Q}q^{-\frac{1+3Q}{4}}\vartheta_{1-Q,0}(2\tau)\,.
\end{equation}
Combining the above equations, we obtain
\begin{equation}\label{eqn:dev2}
\frac{1}{2\pi i}\frac{\partial}{\partial z} ch_{\tilde{R}}^{(h,Q)}(z,\tau)|_{z=0}=\frac{(-1)^{1+Q}}{2\pi}\frac{q^{h-\frac{1+Q}{4}}}{\eta(\tau)^3}\vartheta_{1-Q,0}(2\tau) \vartheta_{1,1}'(\tau)\,.
\end{equation}

Let's now study massless representations. There are three kinds: the vacuum representation $(vac),$ $(+)$, and $(-)$. The vaccum representation has $(h,Q)=(0,0).$ $(+)$ has $(h,Q)=(1/2,1),$ and $(-)$ has $(h,Q)=(1/2,-1).$ The character formulas for the massless representations are obtained by replacing $g(z,\tau;h,Q)$ with
\begin{align}\label{eqn:massless characters}
g^{(vac)}(z,\tau)=& g(z,\tau;0,0)-g\left(z,\tau;\frac{1}{2},1\right)\,,\\
g^{(\pm)}(z,\tau)=&\pm\frac{1}{2}(f_{3,1}(z,\tau)-f_{3,-1}(z,\tau))+\frac{1}{2}g\left(z,\tau;\frac{1}{2},1\right)\,.
\end{align}
Note that the following identities hold
\begin{equation}
f_{3,1}(z,\tau)-f_{3,-1}(z,\tau)=0\,,
\end{equation}
for $z=0,1/2,\tau/2,$ and
\begin{equation}
q^{\frac{3}{8}}\left(f_{3,1}(z,\tau)-f_{3,-1}(z,\tau)\right)=2\,,
\end{equation}
for $z=(\tau+1)/2.$ 
\bibliographystyle{JHEP}
\bibliography{refs}
\end{document}